\documentclass[prl,aps,amsmath,twocolumn,floatfix]{revtex4}

\usepackage{graphics}

\begin{document}



\title{Coupling a single atomic quantum bit to a high finesse optical cavity}
\author{A.~B.~Mundt}
\author{A.~Kreuter}
\author{C.~Becher}
\email{Christoph.Becher@uibk.ac.at}
\homepage{http://heart-c704.uibk.ac.at/Welcome.html}
\author{D.~Leibfried}
\altaffiliation{Present address: Time and Frequency Division, National
Institute of Standards and Technology, Boulder, CO 80305, USA}
\author{J.~Eschner}
\author{F.~Schmidt-Kaler}
\author{R.~Blatt}
\affiliation{Institut f\"ur Experimentalphysik, Universit\"at Innsbruck,
Technikerstra{\ss}e 25, A-6020 Innsbruck, Austria}

\date{\today}


\begin{abstract}

The quadrupole S$_{1/2}$ -- D$_{5/2}$ optical transition of a single trapped
Ca$^+$ ion, well suited for encoding a quantum bit of information, is
coherently coupled to the standing wave field of a high finesse cavity. The
coupling is verified by observing the ion's response to both spatial and
temporal variations of the intracavity field. We also achieve deterministic
coupling of the cavity mode to the ion's vibrational state by selectively
exciting vibrational state-changing transitions and by controlling the
position of the ion in the standing wave field with nanometer-precision.


\end{abstract}


\maketitle


The coherent coupling of a single atom or ion to one mode of the
electromagnetic field inside a high finesse optical resonator is of major
interest for the implementation of quantum information processing schemes:
Single atoms and ions are well suited for storing quantum information in
long-lived internal states, e.g. by encoding a quantum bit (qubit) of
information within the coherent superposition of the S$_{1/2}$ ground state
and the metastable D$_{5/2}$ excited state of Ca$^+$ \cite{NaegerlQubit}. On
the other hand, fast and reliable transport of quantum information over long
distances is most easily achieved by using photons as qubit carriers. The
interface between static and moving qubits is represented by the controlled
interaction of a single atom and a single cavity mode, being the basic
building block for distributed quantum networks \cite{CiracNetwork}.
Deterministic ion-cavity coupling was demonstrated recently by using a single
trapped ion as nanoscopic probe of an optical field
\cite{GuthoehrleinNature}. A second application of atom-cavity coupling
within the field of quantum information processing is the realization of a
deterministic source of single photons
\cite{LawSinglePhotons,HennrichPassage} or sequences of entangled
single-photon wave packets \cite{GheriEntangledPhotons}. More generally,
trapped ions that are cooled to their lowest vibrational state
\cite{Meekhof96,Roos99} and interact with a quantized cavity field might
allow for entangling three quantum subsystems
\cite{BuzekCavityQED,SemiaoVibration}, i.e. internal electronic states,
quantum vibrational mode and single-mode cavity field.
Another application of coupling a trapped ion to a cavity mode is utilizing
the cavity internal standing wave (SW) field \cite{CiracStandWave} or the
cavity-modified spontaneous emission \cite{CiracBadCavity} and coherent
scattering \cite{VuleticCooling} for cooling the ion's vibrational state well
below the Doppler limit.

In this paper we demonstrate coherent coupling of the quadrupole S$_{1/2}$ --
D$_{5/2}$ qubit transition of a single Ca$^+$ ion to a mode of a high finesse
optical cavity. The ion is trapped and placed with high precision at an
arbitrary position in the SW field of the cavity for several hours of
interaction time. We also achieve deterministic coupling of the cavity mode
to the ion's vibrational state by selectively exciting state-changing
transitions with the cavity light tuned to a vibrational sideband of the
S$_{1/2}$ -- D$_{5/2}$ resonance. These demonstrations are important steps
towards realization of the experiments discussed above.

\begin{figure}[h!]
\includegraphics{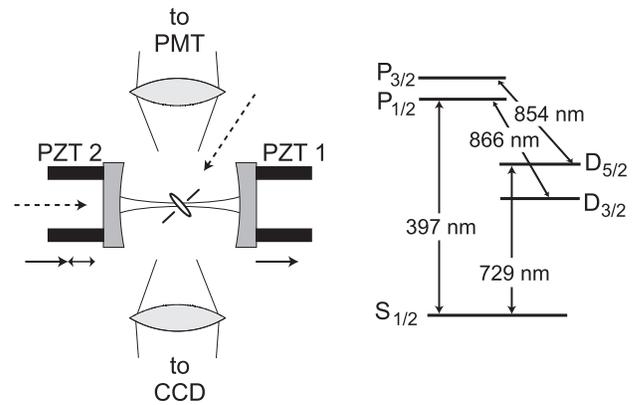}
\caption{\label{setup&level} Schematic experimental setup (left) and Ca$^+$
level scheme (right). PZT1 denotes the offset piezo, PZT2 the scan piezo (see
text). A photomultiplier tube (PMT) is used to record fluorescence on the
S$_{1/2}$ -- P$_{1/2}$ transition and the CCD camera monitors the ion's
position. The following stabilized laser sources are used in the experiment:
two cavity-locked diode lasers at 866~nm and 854~nm with linewidths of
$\approx$~10~kHz and two Ti:Sa lasers at 729~nm ($\approx$~1~kHz linewidth)
and 794~nm ($<$~300~kHz linewidth), where the 794~nm laser is resonantly
frequency doubled to obtain 397~nm. The whole laser system is described in
more detail elsewhere \cite{NaegerlQubit}.The dotted arrows indicate
directions of laser beams: 729~nm excitation laser along the direction of the
cavity axis and 397~nm cooling laser, 854~nm and 866~nm auxiliary lasers at a
certain angle to the trap axis (only shown schematically).}
\end{figure}

The Ca$^+$ ion is stored in a spherical Paul trap \cite{Paul} placed in the
center of a near-confocal resonator. The ion is laser-cooled to the
Lamb-Dicke regime, confining its spatial wave packet to a region much smaller
than the optical wavelength. We detect the coupling of ion and fundamental
TEM$_{00}$ cavity mode by injecting an external light field at 729~nm into
the cavity and recording the excitation on the S$_{1/2}$ -- D$_{5/2}$
transition. The excitation probability is monitored via the electron shelving
technique \cite{Dehmelt75,Roos99}, i.e. by probing the fluorescence on the
S$_{1/2}$ -- P$_{1/2}$ dipole transition (see Fig.\ \ref{setup&level}).


The experimental setup is schematically shown in Fig.\ \ref{setup&level}. The
3-dimensional RF-Paul trap consists of an elliptical ring electrode with
average diameter of 1.4~mm and two endcaps with a spacing of 1.2~mm
(material: 0.2~mm molybdenum wire). The secular frequencies $(\omega_{x},
\omega_{y}, \omega_{z})$ are $2\pi \times (2.9, 3.9, 7.4)$~MHz at a rf drive
field power of 1~W. Here, $z$ denotes the direction of the trap axis, which
is at 45$^{\circ}$ to the cavity axis. The $x$ and $y$ radial directions both
include an angle of $\approx 45^{\circ}$ with the plane spanned by cavity and
trap axis. A magnetic field of 3~G perpendicular to the cavity axis provides
a quantization axis and a frequency splitting of Zeeman components of the
S$_{1/2}$ -- D$_{5/2}$ transition. Calcium ions are loaded into the trap from
a thermal atom beam by a two-step photoionization process using diode lasers
near 423~nm and 390~nm \cite{GuldePhotoion}. The trap is placed in the center
of a near-confocal resonator with finesse ${\cal F} = 35000$ at 729~nm, waist
radius $\omega_0=54~\mu$m, mirror separation $L = 21$~mm and radius of
curvature $R_M=25$~mm. Cylindrical piezoceramics (PZT) allow fine-tuning of
the cavity length across approx. 1.5 free spectral ranges.

The coherent coupling of the ion to the cavity field is measured in three
steps:

(i) {\it Preparation:} First we use Doppler cooling on the S$_{1/2}$ --
P$_{1/2}$ transition at 397~nm (see Fig.\ \ref{setup&level}) to cool the ion
into the Lamb-Dicke regime. A repumper laser at 866~nm inhibits optical
pumping into the D$_{3/2}$ level. From coherent dynamics (Rabi oscillations)
on the carrier and first motional sidebands \cite{Roos99} we determine the
mean vibrational quantum numbers after Doppler cooling to be
$(\bar{n}_{x},\bar{n}_{y},\bar{n}_{z}) = (22.9, 4.3, 4.9)$.  From these mean
phonon numbers and the secular frequencies given above we calculate an rms
extension of the ion's motional wave packet of 26~nm in direction of the
cavity axis, much smaller than the wavelength of 729~nm. After cooling, the
ion is prepared in the S$_{1/2} (m = -1/2)$ substate by optical pumping with
$\sigma^-$ radiation at 397~nm.
%

(ii) {\it Interaction:} The laser at 729~nm is set to a fixed detuning
$\Delta$ from the S$_{1/2}$ -- D$_{5/2}$ ($m=-1/2$ to $m'=-5/2$) qubit
transition. We inject the laser light into the TEM$_{00}$ mode of the cavity
and scan the cavity with a voltage ramp applied to one of the PZTs (scan
PZT). When the cavity reaches resonance with the laser frequency it fills
with light and the ion is excited. A constant voltage is applied to the other
PZT (offset PZT) that determines the ion's position relative to the SW field.

(iii) {\it State analysis:} The final step is state detection by electron
shelving. Fluorescence on the S$_{1/2}$ -- P$_{1/2}$ dipole transition at
397~nm is used to discriminate with high efficiency ($>99\%$) between excited
state (electron shelved in D$_{5/2}$, no fluorescence) and ground state
(fluorescence).

In order to obtain an excitation spectrum, the 729~nm laser is tuned over the
quadrupole transition in steps of about 1~kHz. For any given laser detuning
$\Delta$ the sequence (i)-(iii) is repeated 100 times to determine the
excitation probability.

In our first experiment we probe the ion's response to temporal variations of
the intracavity field by placing it close to a node of the SW field
\cite{quadrupole} and varying the cavity scan rate.
The sign of the voltage ramp applied to the scan PZT determines whether the
scan mirror moves towards the offset mirror or away from it. For a negative
(positive) scan rate, i.e. mirrors moving towards each other (apart), the
intracavity field is Doppler blue (red) shifted and thus the excitation
spectrum will be red (blue) shifted, as the excitation laser detuning has to
compensate for the Doppler shift. The scan velocity is expressed in units of
a normalized scan rate $\nu_L$ \cite{RohdeCavity,LawrenceCavity},
corresponding to the frequency shift in units of HWHM cavity linewidths per
cavity storage time: $\nu_L= 2 {\cal F} \omega \dot{L} \tau_s / \pi c$, with
laser frequency $\omega$, cavity length variation $\dot{L}$ and cavity
(energy) storage time $\tau_s = {\cal F} L / \pi c$.
\begin{figure}[htb]
\includegraphics{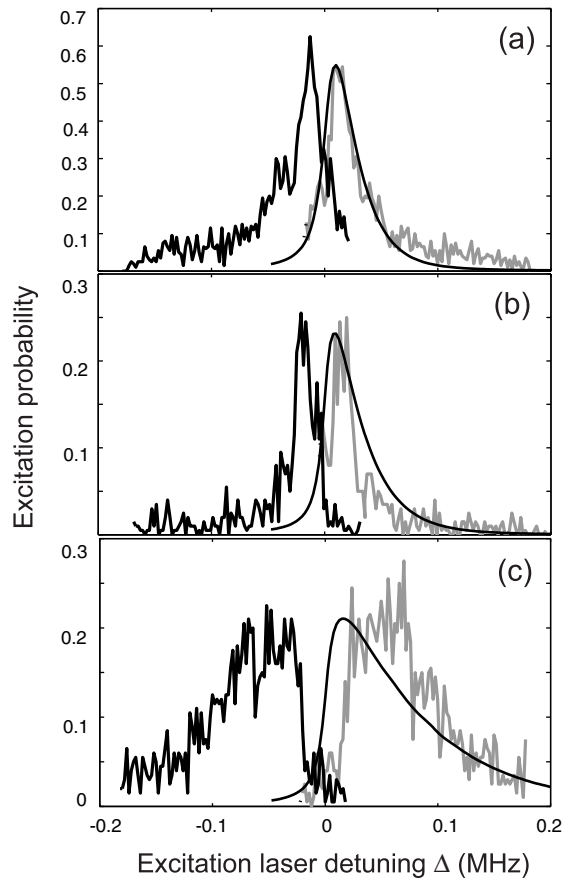}
\caption{\label{scanspec} Excitation spectra of the S$_{1/2}$ -- D$_{5/2}$
transition for different cavity scan rates: $\nu_L= \pm 0.16\ \text{(a)}, \pm
0.23\ \text{(b) and} \pm 0.46\ \text{(c)}$. Blue shifted excitation spectra
are drawn as gray lines on the right hand side of the diagrams, superimposed
solid lines show the theoretical simulation. The parameters used for the
simulations are: excitation laser bandwidth $\Delta \nu_{\text{Laser}}=$
6~kHz, natural linewidth of the S$_{1/2}$ -- D$_{5/2}$ transition $\Delta
\nu_{SD}=$ 0.17~Hz, maximum Rabi frequency at the transition center
wavelength $\Omega_{max}=$ 15.5~kHz (a), 11~kHz (b) and 25~kHz (c), and the
cavity parameters given in the text.}
\end{figure}
The experimental results for scan rates $0.16 > \nu_L
> 0.46$ are shown in Fig.\ \ref{scanspec}.
The excitation spectra show the expected blue shift (red shift) for
increasing positive (negative) scan rates and a broadening due to the Doppler
effect.
An excitation
probability of more than 0.5 as in Fig.\ \ref{scanspec}(a) clearly
demonstrates that the ion is coherently interacting with the intracavity
field.  We theoretically model the excitation for different laser detunings
$\Delta$ by numerically integrating 2-level Bloch-equations using the
time-dependent intracavity field
calculated from the pertaining differential equations
\cite{RohdeCavity,LawrenceCavity}. The results of the theoretical simulation
for positive scan rates are shown as solid lines superimposed on the blue
shifted spectra in Fig.\ \ref{scanspec}. The calculated and experimental
spectra show good agreement for small scan rates (Fig.\ \ref{scanspec}(a,b)).
For larger scan rates (Fig.\ \ref{scanspec}(c)) the centers of the spectra
are slightly shifted. We assume that this shift is caused by nonlinearities
and hysteresis effects of the PZT motion. Although we kept the excitation
power constant in the experiments, we left the Rabi frequency $\Omega_{max}$
as fit parameter to account for variations in excitation due to thermal drift
of the cavity shifting the node of the SW away from the ion's position.

\begin{figure}[h!]
\includegraphics{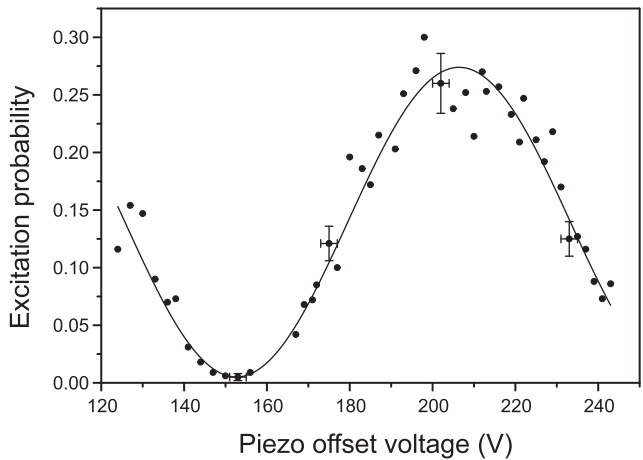}
\caption{\label{standwave} Excitation probability on the S$_{1/2}$ --
D$_{5/2}$ transition of a single trapped Ca$^+$ ion as function of the PZT
offset voltage, i.e. at various positions in the intracavity standing wave
field. The solid line represents a sin$^2$ function fitted to the data
points.}
\end{figure}

The second type of experiment probes the ion's response to spatial field
variations. For this, we leave the scan rate fixed at a small value, allowing
for stable scans with only little perturbations of the excitation spectrum,
as in Fig.\ \ref{scanspec}(a). The intensity of the 729~nm laser is adjusted
such that the excitation is kept well below saturation. The offset voltage of
both scan PZT and offset PZT is then varied simultaneously in such a way that
the SW in the cavity is shifted longitudinally with respect to the location
of the ion. The position-dependent excitation probability is determined by
fitting each excitation spectrum with a Lorentzian and adopting the peak
value. Fig.\ \ref{standwave} displays these values as function of the PZT
offset voltage.
Error bars given for representative data points in Fig.\ \ref{standwave} are
due to PZT hysteresis (abscissa) and the errors of the fits to the excitation
spectra (ordinate). The excitation probability varies spatially with the
intensity of the SW \cite{quadrupole}. A theoretical Bloch-equation analysis
as described above predicts a nearly pure sin$^2$ spatial variation,
deviating by less than 1\%. From a sin$^2$ fit to the data points we obtain
the contrast ratio (visibility $V$) of the position-dependent excitation, $V
= 96.3 \pm 2.6\%$. This very high visibility results from the strong
confinement of the ion's wavefunction: the laser-cooled ion, oscillating with
its secular frequencies and with thermally distributed amplitudes, has a mean
spatial extension along the cavity axis of $a_c = 26$~nm. The thermal
oscillations lead to a calculated reduction of the excitation contrast by a
factor of $\exp(-(2\pi a_c/\lambda)^2) = 95.1\%$, compared to the excitation
contrast experienced by a point-like atom at rest.
Thus, the experimentally determined visibility and the theoretical estimate
agree very well.

A necessary condition for all experiments relying on ion-cavity mode coupling
is the ability to place the ion at a certain position of the intracavity SW
field with high precision and high reproducibility \cite{GuthoehrleinNature}.
In our experiment, the precision of positioning the center of the ion's
wavefunction, using a measurement as in Fig.\ \ref{standwave}, is limited by
the uncertainty in the measured excitation probability. This uncertainty is
due to a statistical error, i.e. the finite number of state detection
measurements, and systematic errors such as fluctuations of laser intensity
and wavelength, drift and jitter of scan PZT etc. \cite{EschnerBarium}. From
the uncertainties in excitation probability (error bars in Fig.\
\ref{standwave}) we deduce a spatial precision between 7~nm ($\approx \lambda
/ 100$) at the position of largest slope and 12~nm and 36~nm at minimum or
maximum excitation, respectively. We note, however, that the precision can be
enhanced by averaging over a larger number of state detection measurements.

Many schemes for quantum information processing with trapped ions rely on
coherent interaction not only with the internal state but also with the
motional degrees of freedom. A controlled coupling to the motional quantum
state is a precondition for realizing such schemes.
\begin{figure}[h!]
\includegraphics{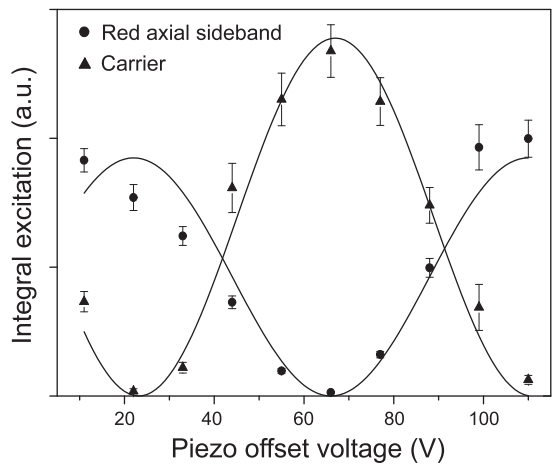}
\caption{\label{standsbcar} Integral excitation on the carrier (triangles)
and the red axial sideband (circles) of the S$_{1/2}$ -- D$_{5/2}$ transition
as function of the PZT offset voltage, i.e. at various positions in the
intracavity standing wave field. The solid lines represent fits of sin$^2$
functions to the data points.}
\end{figure}
In order to demonstrate this motion-dependent coupling, we recorded
excitation probabilities of the ion at a fixed cavity scan rate
($\nu_L=-0.23$), for different positions within the SW, and with the laser at
729~nm now tuned to either the carrier (no change of vibrational quantum
number, $\Delta n =0$) or the red axial sideband ($\Delta n = -1$, laser
detuned by $-\omega_z$) of the S$_{1/2}$ -- D$_{5/2}$ transition. In both
cases, the intensity of the laser was adjusted such that the excitations of
carrier and sideband were comparable and were kept well below saturation. In
this experiment we determine the integral excitation, i.e. the area of the
respective excitation spectra, as the spectra show an asymmetric line shape
(c.f.\ Fig.\ \ref{scanspec}(b,c)). As displayed in Fig.\ \ref{standsbcar},
carrier and sideband excitations both map the SW spatial field variation, but
the traces are shifted by a phase factor of $\pi$. This phase shift arises
due to symmetry characteristics of the transition matrix elements of carrier
and sideband transitions in a SW field \cite{CiracStandWave,SasuraReview}:
The spatial part of the quadrupole transition matrix element is proportional
to $\langle n'|\exp (ikx)|n \rangle$ for a travelling wave (TW) and $\langle
n'|\cos (kx)|n \rangle$ for a SW with the electric field $E \propto \sin(kx)$
\cite{quadrupole}. Here $n$ and $n'$ are the vibrational quantum numbers in
the S$_{1/2}$ and D$_{5/2}$ level, respectively, $k$ is the wavenumber and
$x$ is the ion's position in the field. For a TW, all vibrational states
$(n,n')$ can be coupled as $\exp (ikx)$ contains even and odd powers of $kx$.
In contrary, for a SW $\langle n'|\cos (kx)|n \rangle$ has to be expanded
into even \textit{or} odd powers of $kx$ depending on the ion's position,
e.g. $x=0$ close to a node or $x=\lambda/4$ close to an anti-node. Thus,
transitions changing the phonon number by even or odd integers can only be
excited at different positions in the SW. The red sideband transition
($\Delta n=-1$) couples maximally at anti-nodes of the SW, whereas the
carrier transition ($\Delta n = 0$) couples maximally at nodes.

The high-contrast orthogonal coupling of carrier and sideband transitions to
the cavity mode facilitates applications such as cavity-assisted cooling in a
SW \cite{CiracStandWave} or the Cirac-Zoller quantum-computing scheme
\cite{CiracZoller95}. Both schemes rely on driving vibrational sideband
transitions and benefit from suppressing off-resonant carrier transitions
which induce motional heating or impose a limit on the attainable gate speed
\cite{Steane00}, respectively.
Furthermore, the ion-cavity coupling allows for demonstrating and utilizing
cavity-modified spontaneous emission: 
we calculate the cooperativity parameter $C=g^2 / 2 \kappa \gamma = 0.52$,
 with our experimental ion-field coupling constant $g=2\pi
\times$134~Hz, cavity decay rate $\kappa = 2\pi \times$102~kHz and
spontaneous emission rate $\gamma = 2\pi \Delta \nu_{SD} = 2\pi
\times$~0.17~Hz. Due to the coupling, the spontaneous emission rate is
enhanced by the Purcell factor $F=2C+1=2.04$ and the fraction of spontaneous
emission from the D$_{5/2}$ level emitted into the cavity mode is $\beta =
2C/(2C+1) = 51\%$. This is sufficient for demonstrating cavity-assisted
cooling via destructive quantum interference of heating transitions
\cite{CiracBadCavity}, or triggered single photon emission from the D$_{5/2}$
level. Improvement of the latter scheme and realization of the atom-photon
interface \cite{CiracNetwork} can be readily achieved by employing an
adiabatic transfer technique \cite{LawSinglePhotons,HennrichPassage} and
using a cavity with higher finesse ($\gtrsim 100000$).

In summary we have demonstrated coherent coupling of electronic and motional
states of a single trapped ion to a single field mode of a high finesse
cavity. The position of the ion within the standing wave can be determined
with a precision of up to $\lambda/100$. As the electronic quadrupole
transition in Ca$^+$ is one of the candidates for implementing a quantum bit,
our experiments are a key step towards realization of quantum computing and
communication schemes with trapped ions that require a controlled interaction
of ion and cavity field.


This work is supported by the Austrian 'Fonds zur F\"orderung der
wissenschaftlichen Forschung' (SFB15), by the European Commission (TMR
networks 'QI' and 'QSTRUCT' (ERB-FRMX-CT96-0087 and -0077), IHP network
'QUEST' (HPRN-CT-2000-00121) and IST/FET program 'QUBITS' (IST-1999-13021)),
and by the "Institut f\"ur Quanteninformation GmbH".



\begin{thebibliography}{}
\bibitem{quadrupole}
Note, that the atomic quadrupole moment couples to the gradient of the
electric field. The maximum excitation thus occurs at the nodes of the
standing wave.
\end{thebibliography}
\end{document}